# Coqlex: Generating Formally Verified Lexers

Wendlasida Ouedraogo[a, b], Gabriel Scherer[b, c], and Lutz Straßburger[b, c]

a   Siemens Mobility, Ile-de-France, France
b   INRIA Saclay, Ile-de-France, France
c   Laboratoire d'Informatique de l'École polytechnique, Ile-de-France, France

**Abstract**
**Context**   A compiler consists of a sequence of phases going from lexical analysis to code generation. Ideally, the formal verification of a compiler should include the formal verification of each component of the tool-chain. An example is the CompCert project, a formally verified C compiler, that comes with associated tools and proofs that allow to formally verify most of those components.
**Inquiry**   However, some components, in particular the lexer, remain unverified. In fact, the lexer of Compcert is generated using OCamllex, a lex-like OCaml lexer generator that produces lexers from a set of regular expressions with associated semantic actions. Even though there exist various approaches, like CakeML or Verbatim++, to write verified lexers, they all have only limited practical applicability.
**Approach**   In order to contribute to the end-to-end verification of compilers, we implemented a generator of verified lexers whose usage is similar to OCamllex. Our software, called Coqlex, reads a lexer specification and generates a lexer equipped with a Coq proof of its correctness. It provides a formally verified implementation of most features of standard, unverified lexer generators.
**Knowledge**   The conclusions of our work are two-fold. Firstly, verified lexers gain to follow a user experience similar to lex/flex or OCamllex, with a domain-specific syntax to write lexers comfortably. This introduces a small gap between the written artifact and the verified lexer, but our design minimizes this gap and makes it practical to review the generated lexer. The user remains able to prove further properties of their lexer. Secondly, it is possible to combine simplicity and decent performance. Our implementation approach that uses Brzozowski derivatives is noticeably simpler than the previous work in Verbatim++ that tries to generate a deterministic finite automaton (DFA) ahead of time, and it is also noticeably faster thanks to careful design choices.
**Grounding**   We wrote several example lexers that suggest that the convenience of using Coqlex is close to that of standard verified generators, in particular, OCamllex. We used Coqlex in an industrial project to implement a verified lexer of Ada. This lexer is part of a tool to optimize safety-critical programs, some of which are very large. This experience confirmed that Coqlex is usable in practice, and in particular that its performance is good enough. Finally, we performed detailed performance comparisons between Coqlex, OCamllex, and Verbatim++. Verbatim++ is the state-of-the-art tool for verified lexers in Coq, and the performance of its lexer was carefully optimized in previous work by Egolf and al. (2022). Our results suggest that Coqlex is two orders of magnitude slower than OCamllex, but two orders of magnitude faster than Verbatim++.
**Importance**   Verified compilers and other language-processing tools are becoming important tools for safety-critical or security-critical applications. They provide trust and replace more costly approaches to certification, such as manually reading the generated code. Verified lexers are a missing piece in several Coq-based verified compilers today. Coqlex comes with safety guarantees, and thus shows that it is possible to build formally verified front-ends.

## The Art, Science, and Engineering of Programming

**Coqlex: Generating Formally Verified Lexers**

# 1 Introduction

A lexer is a tool that is in charge of the lexical analysis, one of the first phases of compilers and interpreters. Lexers take a sequence of characters (such as source code or command) as input and produce a sequence of tokens (parts of that input sequence of characters associated with meaning) that can be easily processed by parsers. During that process, lexers can ignore comments or white spaces, and also equip tokens with source position information (such as line numbers) to enable useful debug messages during lexical analysis (lexing), parsing, or later compilation stages.

Implementing a lexer from scratch can be difficult and time-consuming. This has led researchers to build tools, libraries, and generators to help to implement optimized lexers. Most of the existing implementations of those libraries and generators, such as OCamllex [20], do not come with formal proof of correctness.

This is the starting point for our work on *Coqlex*,[1] a generator of formally verified lexers. Our goal was to provide a tool that is as versatile as OCamllex and that at the same time is formally verified, so that it can be integrated into formally verified compiler tool-chains, such as CompCert [14].

The main issues with the formal verification of tools such as lexers are related to (i) the execution time, (ii) the integration with existing parsers, and (iii) the usability. This document presents techniques we used to tackle those challenges. Our contributions are as follows:

1. **The verification of lexical rule selection.** Most lexer generators produce lexers from lexer specification files. Those specifications define a lexer using *lexing rules* that are pairs of input patterns, defined via regular expressions [10, 22] (regexp), and semantic actions that are in charge of production tokens. Depending on the selection policy and the text to analyse, the lexer selects a pair by analysing its regexps. When a pair is selected, the token to produce is handled by its semantic action. Coqlex implements two selection policies (the longest match and the shortest match associated with the priority rules) and provides a Coq proof of their correctness.

2. **The *Coqlex generator*.** Coqlex also provides a small preprocessor (the *Coqlex generator*) that lets users specify lexers in user-friendly syntax, inspired by the one of OCamllex. There is no simple specification for this input syntax itself. Instead, the *Coqlex generator* translates it to a human-readable Coq file with (heavier notation but) the same structure, and the correctness statement is given in terms of this translated source. This is similar to the "Coq production" mode of the Menhir parser generator [9].

In summary, a user of Coqlex writes a lexer in a familiar syntax close to standard lexer generators, and gets a human-readable description of the lexing rules in Coq. They should review the lexing rules in Coq syntax to ensure that it corresponds to their intent. Coqlex then provides a function that takes those lexing rules and interprets them into a lexing function, with a formal proof that this function is correct.[2]

---

[1] the source code of *Coqlex* can be found on https://gitlab.inria.fr/wouedrao/coqlex
[2] Our lexer functions are correct with respect to a standard specification discussed in Section 5.

After having finished this work, we discovered the independent research on the Verbatim++ [6, 7] formally verified lexer. For this reason, we will here also include:

3. **A comparison with other lexer generators.** More precisely, we compare Coqlex to OCamllex and flex (because these are standard tools) and to Verbatim++ (the state of the art for verified lexers in Coq), with respect to execution time and usability.

In OCamllex, lexical rules are compiled into a non-deterministic automaton represented by a compact table of transitions, with backtracking and semantic actions. The generated lexer code simply follows the automaton transition by repeated table lookups. In contrast, Coqlex contains no such compilation, it interprets the user-provided regular expressions against the input by using Brzozowski derivatives [4]. This allows to have a simpler formalization, and leads to surprisingly good performance: roughly 100x slower than OCamllex, but 100x faster than Verbatim++. This is reasonable for a pure program extracted directly from Coq, compared to an efficient implementation, and is more than fast enough in practice. For example, the lexer of the *Coqlex generator* is implemented in Coq, using the Coqlex data structures and functions. That generator is used without any noticeable slowness.

The Verbatim++ lexer (also verified in Coq) implements regexps using Brzozowski derivatives and then compiles those regexps into deterministic finite automata[3] for fast regexp matching. Even though significant emphasis in the work [6, 7] is put on optimization, Verbatim++ remains substantially slower than Coqlex in our experiments (which used the benchmarks provided by Verbatim++).

Our work highlights the fact that a simple model and implementation can lead to good enough performance. It also provides a complete lexer generator (an executable) and library that allow to implement lexers easily in Coq and then extract them into OCaml code. Those lexers, written in Coq, come with proven lemmas that allow developers to prove specific properties on them. In addition, associated with Menhir [19] verified parsers, Coqlex verified lexers allow to write fully formally verified front-ends for formally verified compilers such as CompCert [14].

This paper is organized as follows: In Section 2 we discuss the representation of a lexer in Coq. In Section 3 we present the *Coqlex generator* and present our industrial use-case for Coqlex in Section 4. We discuss its specification and correctness in Section 5. Section 6 presents the implementation details of Coqlex. Section 7 compares the features and performance of Coqlex with OCamllex and Verbatim++. Finally, we discuss future work and conclude in Section 9.

## 2 Representing a Lexer in Coq

From a functional point of view, lexers are in charge of producing tokens (user-defined type that we will note T) from a text (string). A natural type would be

lexer(T) := string -> list T

Lexers like those generated by most lexer generators, in particular OCamllex[20], produce tokens one by one and are called by parsers on demand. A function that performs *one step* of lexical analysis (lexing) consumes a string and returns a token and the remaining string. The type of such a function is

    lex1(T) := string -> T * string

Lexing can fail for various reasons. In case of failure, lexers should provide useful error messages. For that reason, we defined a position data type and an error data type to encapsulate the lexing result (Result). A function that performs one step of lexing becomes a function that takes an input string, a start position, and in case of success, returns a token, the remaining string, and the end position. Consequently, the type of one step of lexing becomes

    lex1(T) :=   string ->
                   position ->
                   Result(T * string * position)

Most lexer generators generate lexers using a set of lexical rules that are regular expressions[10, 22] (regexp) associated with semantic actions that are in charge of producing the lexing result. The semantic action that will produce the returned lexing result is the first one associated with the regexp that matches the longest prefix of the input string (the lexeme): this is the longest match and the priority rules. Semantic actions have access to the lexing buffer (lexbuf), a data structure containing the lexeme, the start position (the position of the first letter of the lexeme), the end position (the position of the letter after the last letter of the lexeme) and the remaining string (the input string without the lexeme). The semantic action also specifies how the internal state of the lexer should be updated. So, a natural type for semantic actions would be:

    action(T) :=   lexbuf ->
                     Result(T * string * position)

Those semantic actions can perform various operations, including recursive calls to the lexer that calls them. This could then lead to an infinite loop.[3] As Coq forbids the implementation of functions that loop[5], Coqlex had to find a solution to deal with those kinds of situations. We explored three possibilities:

1. Making restrictions on semantic actions that ensure termination. For example, we could require that each semantic action discards at least one character from the input string.
2. Letting developers write their own termination proofs for all the lexer they generate.
3. Using the fuel technique: this technique consists of ensuring the termination of lexers using a natural number (nat) that decreases at every recursive call.

Requiring that each semantic action discards at least one input character is too strict in practice. Studying lexers in the wild, we have found many cases of lexers designed to "skip" an optional part of the input, that accept the empty string if nothing needs to be skipped. For example, the lexer of the OCaml compiler contains the following:

---

[3] Section 7.3 provides a typical OCamllex example of a lexer that can loop due to recursive calls.

```
1  rule skip_hash_bang = parse
2  | "#!" [^ '\n']* '\n' { new_line lexbuf }
3  | "" { () } (* accepts the empty string *)
```

Forcing users to write termination proofs for potentially non-terminating lexers would force users to review and complete the generated Coq code and so alter the user experience.

We thus chose to express general, potentially non-terminating lexers using fuel. Consequently, the type of one step of lexing becomes

$$\begin{aligned} \text{lex1(T)} := \quad & \text{nat ->} \\ & \text{lexbuf ->} \\ & \text{Result(T * string * position)} \end{aligned}$$

To make it simple to call a lexer from a semantic action, we replace the separate arguments string and position by the more informative lexbuf type already used with semantic actions.

$$\begin{aligned} \text{lex1(T)} := \quad & \text{nat ->} \\ & \text{action(T)} \end{aligned}$$

$$\begin{aligned} \text{action(T)} := \quad & \text{lexbuf ->} \\ & \text{Result(T * lexbuf)} \end{aligned}$$

## 3 Coqlex in Practice

Coqlex comes with a Coq library that allows to write lexers using sets of lexical rules. It also provides a text processor that will convert a markup language (.vl syntax), which is similar to the OCamllex[20] specification language (.mll syntax), into its equivalent Coq code (.v file). Figure 1 presents the .vl version of the mini-cal (a micro language for arithmetic expressions: numbers, idents, + * - / and parentheses) lexer and its equivalent in .mll. This .vl definition has four parts:

1. The header section: The header section is arbitrary Coq text enclosed in curly braces. If present, the header text is copied as it is at the beginning of the output file. Typically, the header section contains the Coq Require Import directives, possibly some auxiliary functions, and token definitions used for lexer definitions.
2. The regexp definition section: This section allows to give names to frequently-occurring regular expressions. This is done using the syntax let *ident* = *re* to associate the name *ident* to the regexp *re*. The syntax of regexp is defined in Figure 2.
3. The lexer definition section: This section allows to define lexers using sets of rules. A rule is defined using the syntax | *p* {*a*} (the '|' symbol is not mandatory for the first rule) to associate the pattern *p* to the Coq text representing a semantic action *a*. This pattern is either a regexp or a string -> bool function (defined using the syntax $(f)$ where $f$ is the Coq code of this function). Typically, this kind of pattern is

```
1  (**Coqlex**)
2
3  (* header section *)
4  {
5  Require Import TokenDefinition.
6
7  }
8
9  (* regexp definitions *)
10 let ident = ['a'-'z']+
11 let numb = ['0'-'9']+
12
13 (* lexer definitions*)
14 rule minlexer = parse
15 | '\n' { sequence [new_line; minlexer] }
16 | ident {ret_l ID}
17 | numb { ret_l Number }
18 | '+' { ret PLUS }
19 | '-' { ret MINUS }
20 | '*' { ret TIMES }
21 | '(' { ret LPAREN }
22 | ')' { ret RPAREN }
23 | eof { ret Eof }
24 | _ { raise_l "unknown token :"}
25
26 (* trailer section *)
27 {}
```

```
1  (**OCaml**)
2
3  (* header section *)
4  {
5  open Lexing
6  open TokenDefinition
7  }
8
9  (* regexp definitions *)
10 let ident = ['a'-'z']+
11 let numb = ['0'-'9']+
12
13 (* lexer definitions*)
14 rule minlexer = parse
15 | '\n' { new_line lexbuf; minlexer lexbuf }
16 | ident {ID (Lexing.lexeme lexbuf)}
17 | numb { Number (Lexing.lexeme lexbuf)}
18 | '+' { PLUS }
19 | '-' { MINUS }
20 | '*' { TIMES }
21 | '(' { LPAREN }
22 | ')' { RPAREN }
23 | eof { Eof }
24 | _ { failwith ("unknown token : " ^ (Lexing.lexeme lexbuf))}
25
26 (* trailer section *)
27 {}
```

**Figure 1** Comparing Coqlex and OCamllex mini-cal lexer definition

used to detect situations in which the lexing must stop (e.g when the input string is empty). When the pattern is a regexp, the semantic rule is said to be regexp based. Otherwise, the semantic rule is said to be function based.

4. The trailer section: This section is similar to the header section, except that its text is copied as it is at the end of the output file. Typically, this section contains Coq extraction directives.

**Remarks:**
- A .vl file allows to define multiple lexers. Those lexers are gathered in groups (made of mutually recursive lexers) using the keyword *and*. To define non-mutually recursive lexers, the user should use the keyword *then* instead.
- *Coqlex generator* users do not need to worry about the management of fuel when writing .vl files: for a given input fuel $n$, the fuel of the lexers that are called in semantic actions are either $n-1$ (for recursive calls) and $n$ otherwise.
- The default starting fuel constant we use is 1 000 000. This arbitrary large value was enough for our tests. This bounds the depth of recursive calls inside semantic actions for one token, rather than the number of calls for the whole input.
- Developers can change the value of the starting fuel using the –fuel switch.

$re ::=$
   'c'                 Character constant
   | "$string$"      String constant
   | _                 Char wildcard
   | $[s_1 s_2 ... s_n]$     Union of character sets
   | $[\char`\^ s_1 s_2 ... s_n]$    Negation of union of character sets
   | $re_1 | re_2$       Alternative
   | $re_1\ re_2$        Concatenation
   | $re_1 - re_2$       Difference of regexps.
                         It accepts a string that is accepted by $re_1$ and rejected by $re_2$
   | $re*$             Kleene star
   | $re+$             Strict repetition
   | $re?$             Option

$s ::=$
   'c'               Character constant
   | '$c_1$' - '$c_2$'    Character range

■ **Figure 2** Syntax of Coqlex regexps

- When a lexer loops for a given input string, it will always run out of fuel for that input string, regardless of the input fuel. However, a lexer running out of fuel for a given input does not necessary mean that this lexer loops.

In Figure 3, we compare the description mini-cal lexer in Coqlex syntax to the code it generates. Typically, the generator translates the regexp written in .vl syntax into the Coqlex regexp data type. The generated lexing function also calls Coqlex functions such as generalizing_elector, longest_match_elector, and exec_sem_action. It also recurses over the fuel explicitly; we could instead generate a call to a fixpoint combinator, but this would be difficult to scale to mutually-recursive lexers.

Similarly to OCamllex, Coqlex also allows to choose the semantic action by matching the shortest prefix. In that case, the user should use the keyword shortest instead of parse in the .vl file. In the generated file, the function longest_match_elector is then replaced by the function shortest_match_elector.

## 4 Coqlex Industrial Use-case

The development of Coqlex finds its origin in an industrial project that aims at optimizing programs that are executed on a DIGISAFE® XME vital computer[13]. Those programs are safety-critical, follow strict certification processes (SIL-4), and are built by a long chain of transformations from high-level models to executable code. One part of the transformation pipeline is a source-to-source optimizer, at a level where programs are represented in Ada (generated by model-driven engineering tools

**Coqlex: Generating Formally Verified Lexers**

```
 1  (**Coqlex input**)
 2  
 3  { Require Import TokenDefinition. }
 4  
 5  let ident = ['a'-'z']+
 6  
 7  
 8  
 9  let numb = ['0'-'9']+
10  
11  
12  
13  
14  rule minlexer = parse
15  | '\n' { sequence [new_line; minlexer] }
16  | ident {ret_l ID}
17  | numb { ret_l Number }
18  | '+' { ret PLUS }
19  | '-' { ret MINUS }
20  | '*' { ret TIMES }
21  | '(' { ret LPAREN }
22  | ')' { ret RPAREN }
23  | eof { ret Eof }
24  | _ { raise_l "unknown token :"}
```

```
 1  (**Coqlex output**)
 2  
 3  Require Import TokenDefinition.
 4  
 5  Definition ident := RegexpSimpl.simp_cat
 6   ((RValues.const_CharRange "a"%char  "z"%char ))
 7   (RegexpSimpl.simp_star
 8    ((RValues.const_CharRange "a"%char  "z"%char ))).
 9  Definition numb := RegexpSimpl.simp_cat
10   ((RValues.const_CharRange "0"%char  "9"%char ))
11   (RegexpSimpl.simp_star
12    ((RValues.const_CharRange "0"%char  "9"%char ))).
13  
14  Fixpoint minlexer  fuel  lexbuf {struct fuel} := match fuel with
15  | 0 => (AnalysisNoFuel lexbuf)
16  | S n => ( match LexerDefinition.generalizing_elector
17      LexerDefinition.longest_match_elector ([
18      (Char "010"%char ,  sequence [new_line; (minlexer n)] );
19      (ident, ret_l ID);
20      (numb,  ret_l Number );
21      (Char "+"%char ,  ret PLUS );
22      (Char "-"%char ,  ret MINUS );
23      (Char "*"%char ,  ret TIMES );
24      (Char "("%char ,  ret LPAREN );
25      (Char ")"%char ,  ret RPAREN );
26      (RValues.regex_any,  raise_l "unknown token : ")],
27      [(CoqLexUtils.EOF,  ret Eof )]
28    ) (remaining_str lexbuf) with
29    | Some elt => LexerDefinition.exec_sem_action elt lexbuf
30    | None => (AnalysisFailedEmptyToken lexbuf)
31    end)
32  end.
```

■ **Figure 3** Comparing Coqlex generator input and output for mini-cal language in the VCP Ada subset). We worked on verifying this source-to-source optimizer, with verified optimizations but also a verified frontend using Coqlex (lexer) and Menhir with Coq back-end (parser).

Our Coqlex lexer for Ada is written in 200 lines of code. The source-to-source compiler was tested on projects of various sizes; the largest project we compiled has 2380 files totalling 25MB of VCP Ada (a subset of Ada) code – the VCP Ada code is not written by humans as-is but generated by other tools in the pipeline, so it can grow very large.

This use-case motivated several changes to Coqlex. The first version of Coqlex imposed termination proofs but any change in the lexer could require rewriting those proofs. Another version imposed restrictions on semantic actions and defined lexers as records that contain the lexer function and properties that ensures termination, making lexer description non-trivial, and mutual recursion impossible. Another version imposed an additional input and output parameter that allows lexers to communicate during mutually recursive calls, typically for saving comments content.

## 5 *Coqlex Generator* Specification

Given a set of regexp-based rules $l_r$, a set of function-based rules $l_f$, and a matching policy $e$, the generated Coq code implements a lexer — a function that takes a fuel $n_f$, lexbuf $b$, a storage $s$ and returns a lexing result — that works as follows:

1. If $n_f$ is equal to 0, then the result is an error. This error is a direct consequence of the fuel technique.
2. Otherwise, from the input string, $l_r$, $l_f$ and $e$, the lexer chooses a rule whose semantic action will be in charge of returning the lexing result.
   a. if the selected rule is a function-based one, made up with a function $f$ associated with a semantic action $a$, then there is no consumption. Consequently, the lexing result is the result of $a$ called with $b$ and $s$.
   b. if the selected rule is a regexp-based rule $c$ – made up with a regexp $r$ associated with a semantic action $a$ – and if the length of the prefix matched by $r$ using the policy $e$ is a natural number $n$, then the lexing result is the result of $a$ applied with the updated lexbuf $b_u$ and the input storage $s$. The updated lexbuf is defined as follows:
      - the lexeme of $b_u$ is the $n$ first characters of the input string.
      - the remaining string of $b_u$ is the input string without the lexeme.
      - the end position of $b_u$ is the end position of $b$ where the column number is incremented by $n$.
      - the start position of $b_u$ is the end position of $b$.
   c. if no rule is selected, the lexer must return an error meaning that the input string contains elements that cannot be analysed by the lexer.

Except for the use of fuel, the functioning of the generated lexer defined above is standard.

A .vl file provides the description of lexers using lexical rules. That description is processed by the *Coqlex lexer generator* whose architecture is detailed in Figure 4. It has three components:

1. The lexer, that is in charge of generating a set of tokens from the text of the .vl file, is written in Coq using the Coqlex library and is formally verified.
2. The parser, that is in charge of generating an abstract representation from the set of tokens produced by the lexer, is implemented using Menhir [19] with `-coq` switch to generate verified parsers.
3. The code printer, that is in charge of generating the .v file from the abstract representation produced by the parser, is written in OCaml and is not formally verified.

The code printer does not include formal semantics equivalence between the representation of the .vl code and the generated .v code. This means that, *a priori*, a critical user should review the generated .v code. This does not take great efforts because the transformation does not include a complex compilation process: the .vl and .v files have similar structures and are human readable.

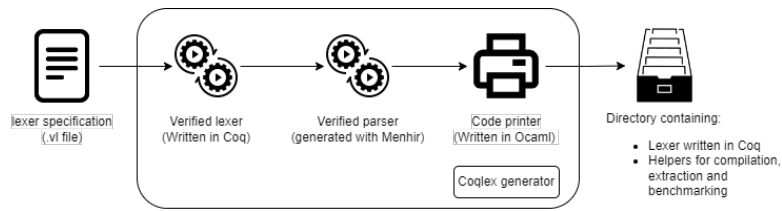

**Figure 4** General structure of the *Coqlex lexer generator*.

By comparison, Verbatim++ does not provide such a generation tool, and OCamllex generates an OCaml code in which the patterns of the lexical rules are compiled into a non deterministic automaton[3] represented by a compact table of transitions, making the generated code non human readable. Figure 18 in Appendix C compares the OCamllex code with its generated lexer for mini-cal.

In our case, the regexps, rule selection, and associated policies used in the generated file are implemented and proved correct in Coq. Consequently, the attention of the critical user who wants to check the generated .v file must be focused on the following elements:

- The translation of regexps: The user must be assured of the correspondence between the regexps written in the input .vl files and those generated in the output .v files. This requires to read and understand the regexps constructors that will be defined in Section 6.
- The matching policy: The user has to make sure that the matching policy corresponds to the one that is described in the .vl file. The keyword parse must correspond to longest_match_elector and shortest must correspond to shortest_match_elector.
- For every lexer, the user must be assured that the right regexps are associated with the right semantic actions and in the same order. In the Coq code, a difference is made between lexical rules made up with regexps associated with semantic actions (regexp-based rules) and those made up with string -> bool functions associated with semantic actions (function-based rules).

In a nutshell, a potential user has to (i) review the Coq implementation and verification of regexps, the rule selection together with the associated policies and helpers, that are written and proved in Coq, once; and (ii) either review the code printer of the *Coqlex generator* once, or review the elements listed above at every generation.

## 6 Coqlex Implementation Details

Most lexer generators such as OCamllex speed up lexical analysis by compiling lexical rules into finite automata during lexer generation. In Coqlex, lexical rules are interpreted on the fly, using Brzozowski derivatives[4] for regexps and simple functions for matching policies.

$regex ::=$
  $\emptyset_r$    The empty regexp
       $L(\emptyset_r) = \emptyset$
  $|\ \epsilon_r$    The empty string regexp
       $L(\epsilon_r) = \{\epsilon\}$
  $|\ [\![a]\!]$    The one-symbol regexp ($a \in \mathbb{A}$)
       $L([\![a]\!]) = \{a\}$
  $|\ e_1 + e_2$    The alternative
       $L(e_1 + e_2) = L(e_1) \cup L(e_1)$
  $|\ e_1 \cdot e_2$    The concatenation
       $L(e_1 \cdot e_2) = L(e_1) \mathbin{+\!\!+} L(e_2)$
  $|\ e^*$    The Kleene star
       $L(e^*) = \bigcup_{n \in \mathbb{N}} L(e)^n$

■ **Figure 5** Definition of regular expressions associated with the language they describe. Variables $e$, $e_1$ and $e_2$ are regular expression. The symbol $\{a\}$ denotes the set containing a unique string that is made up with a unique symbol which is $a$.

| | |
|---|---|
| nullable $\emptyset_r$ = false | nullable $\epsilon_r$ = true |
| nullable $[\![a]\!]$ = false | nullable $(e_1 + e_2)$ = nullable $e_1$ ∨ nullable $e_2$ |
| nullable $(e_1 \cdot e_2)$ = nullable $e_1$ ∧ nullable $e_2$ | nullable $e^*$ = true |

■ **Figure 6** Definition of the nullable function. The variable $a$ stands for a symbol and variables $e$, $e_1$ and $e_2$ for regular expressions.

### 6.1 Brzozowski Derivatives for Regexps Matching

Given an alphabet (set of symbols or characters) $\mathbb{A}$, we use the symbol $\epsilon$ to denote the empty string, and we use the operator $+\!\!+$ for string concatenation. This operation can be extended to languages via $L_1 \mathbin{+\!\!+} L_2 = \{s_1 \mathbin{+\!\!+} s_2 \mid s_1 \in L_1 \wedge s_2 \in L_2\}$. Then we define $L^0 = \{\epsilon\}$ and $L^{n+1} = L \mathbin{+\!\!+} L^n$ (for $n \in \mathbb{N}$). Finally, we use the notation $L(r)$ to denote the language of the regexp $r$, which is inductively defined in Figure 5.

Using all the notations above, we say that a regular expression $r$ matches a string $s$ if $s \in L(r)$. Similarly, when $s \notin L(r)$ we say that $r$ does not match $s$.

Coqlex uses regexp constructions and matching algorithms based on the concept of Brzozowski derivatives. This concept introduces two functions: the nullable function and the derivative of a regexp.

The nullable function takes a regexp $r$ and returns the boolean true if $r$ matches $\epsilon$ (the empty string) and false otherwise. Its inductive definition is given in Figure 6.

The derivative of a regexp does match the derivative of the corresponding language. More precisely, if $az$ denotes the string built from the symbol $a$ as first element and the string $z$, then the derivative of a regular expression $r$ by a symbol $a$ is the regexp

$$
\begin{aligned}
\varnothing_r/c &= \varnothing_r \\
\epsilon_r/c &= \varnothing_r \\
[\![a]\!]\;/\;c &= \begin{cases} \epsilon \text{ if } a = c \\ \varnothing_r \text{ otherwise} \end{cases} \\
(e_1 + e_2)/c &= (e_1/c) + (e_2/c) \\
(e_1 \cdot e_2)/c &= \begin{cases} (e_1/c \cdot e_2) + e_2/c \text{ if nullable } e_1 = \text{true} \\ (e_1/c \cdot e_2) \text{ otherwise} \end{cases} \\
e^*/c &= (e/c) \cdot e^*
\end{aligned}
$$

■ **Figure 7** Definition of the derivative of a regexp. The variables $a$ and $c$ stand for symbols and variables $e$, $e_1$ and $e_2$ for regular expression.

$$
\begin{aligned}
r \| \epsilon &= r \\
r \| az &= (r/a) \| z
\end{aligned}
$$

■ **Figure 8** Extension of the derivative of a regexp to strings. Variables $r$, $\epsilon$, $a$ and $z$ denote, respectively, a regex, the empty string, a symbol and a string. The operator / refers to the derivative operation described in Figure 7. The notation $az$ denotes the string composed of the symbol $a$ as first element and the string $z$.

$r/a$ that denotes the language $\{z \mid az \in L(r)\}$. The inductive definition is given in Figure 7.

Brzozowski [4] extended the derivative operation to strings (denoted by $\|$) as described in Figure 8, and showed that for every regular expression $r$ and every string $s$

$$s \in L(r) \iff \text{nullable } (r \| s) = \text{true}$$

Coqlex uses a modified version of an existing Coq implementation[16] of Brzozowski derivatives for regexp matching. That implementation comes with safety guarantees as it provides a Coq proof showing that this Brzozowski derivative implementation is a Kleene algebra[11] and provides common proofs on regexps. It also provides equivalence ($\equiv$) relation on regexps and additional regexp constructors such as the conjunction and negation constructors that are not used in the regexp constructors that are provided by the *Coqlex generator* (see Figure 2). On the other hand, some of the constructions of regexps presented in Figure 2 are missing. For this reason, we modified the existing Coq implementation [16] of Brzozowski derivatives as follows:

1. We removed the conjunction and negation regexp constructors
2. We added five regexp constructors:

- **the negation of a symbol:** The notation $\overline{[\![a]\!]}$ denotes a regexp that matches any 1-length-string whose character is not equal to $a$. This regexp is defined by the following two properties: nullable $\overline{[\![a]\!]}$ = false and for all symbol $s$,

$$\overline{[\![a]\!]}/c = \begin{cases} \epsilon_r & \text{if } c \neq a \\ \varnothing_r & \text{otherwise} \end{cases}$$

- **the char wildcard:** The notation $\omega_r$ denotes a regexp that matches any 1-length-string. This regexp is defined by the following two properties: nullable $\omega_r$ = false and for all symbol $s$, $\omega_r/c = \epsilon$. Then, we proved that for all strings $s$, we have $s \in L(\omega_r)$ if and only if $s$ consists of a single character.

- **the character set:** The notation $\Sigma_l^u$ (where $l$ and $u$ are symbols) denotes a regexp whose language is $L(\Sigma_l^u) = \{c | l \leq c \wedge c \leq u \wedge c \in \mathbb{A}\}$ (where $\leq$ is a reflexive, antisymmetric and transitive order relation on symbols). This constructor is defined using the following two properties: nullable $\Sigma_l^u$ = false and for all symbols $c$

$$\Sigma_l^u/c = \begin{cases} \epsilon_r & \text{if } l \leq c \wedge c \leq u \\ \varnothing_r & \text{otherwise} \end{cases}$$

We proved that if $\neg(l \leq u)$, then $\Sigma_l^u \equiv \varnothing_r$ and that for every string $s$, $s \in L(\Sigma_l^u)$ if and only if $s$ consists of only one symbol $c$ such that $l \leq c \wedge c \leq u$.

- **the negation of character set:** The notation $\overline{\Sigma_l^u}$ (where $l$ and $u$ are symbols) denotes a regexp whose language is $L(\overline{\Sigma_l^u}) = \{c | \neg(l \leq c \wedge c \leq u) \wedge c \in \mathbb{A}\}$. This constructor is defined using the following two properties: nullable $\overline{\Sigma_l^u}$ = false and for all symbols $c$

$$\overline{\Sigma_l^u}/c = \begin{cases} \epsilon_r & \text{if } \neg(l \leq c \wedge c \leq u) \\ \varnothing_r & \text{otherwise} \end{cases}$$

We proved that if $\neg(l \leq u)$, then $\overline{\Sigma_l^u} \equiv \omega_r$ and that for every string $s$, $s \in L(\overline{\Sigma_l^u})$ if and only if $s$ consists of only one symbol $c$ such that $\neg(l \leq c \wedge c \leq u)$.

- **the difference:** The notation $e_1 - e_2$ (where $e_1$ and $e_2$ are regexps) denotes a regexp whose language is $L(e_1 - e_2) = \{s | s \in L(e_1) \wedge s \notin L(e_2)\}$. This construction is defined using the following two properties: nullable $e_1 - e_2$ = (nullable $e_1$) $\wedge \neg$(nullable $e_2$) and for all symbol $c$, $(e_1 - e_2)/c = e_1/c - e_2/c$. We proved that for all strings $s$, we have $s \in L(e_1 - e_2) \iff s \in L(e_1) \wedge s \notin L(e_2)$.

These constructors have also been added for performance reasons. In fact, the regexp $\Sigma_{c_n}^{c_{n+m}}$ could be written as $[\![c_n]\!] + [\![c_{n+1}]\!] + ... + [\![c_{n+m}]\!]$. However, using the first representation ($\Sigma_{c_n}^{c_{n+m}}$), the derivation function will perform 2 comparisons, while the second one will perform $m + 1$ comparisons (see Figure 7).

### 6.2 Matching Policies

Coqlex defines two types of rules: function based and regexp based ones. During the lexical analysis, the generated lexer has to select a rule. This selection starts with the

**Coqlex: Generating Formally Verified Lexers**

$$\overline{E_f([],s) = \bot} \qquad \frac{f\ s = \text{true}}{E_f((f,a) :: t, s) = (f, a)} \qquad \frac{f\ s = \text{false}}{E_f((f,a) :: t, s) = E_f(t, s)}$$

■ **Figure 9** The formal description of the selection of a function based-rule. This description uses the list notation: [] denotes the empty list and $h :: t$ denotes a list whose first element is $h$ and whose tail is $t$. The symbol $\bot$ means that no rule is selected.

$$\frac{\text{nullable } r = \text{true}}{\mathbb{S}_l(r, \epsilon) = 0} \qquad \frac{\text{nullable } r = \text{false}}{\mathbb{S}_l(r, \epsilon) = -\infty} \qquad \frac{\mathbb{S}_l(r/a, z) = n}{\mathbb{S}_l(r, az) = n + 1}$$

$$\frac{\mathbb{S}_l(r/a, z) = -\infty \quad \text{nullable } r = \text{true}}{\mathbb{S}_l(r, az) = 0} \qquad \frac{\mathbb{S}_l(r/a, z) = -\infty \quad \text{nullable } r = \text{false}}{\mathbb{S}_l(r, az) = -\infty}$$

■ **Figure 10** The formal description of *l-score* computation.

choice of a function based rule (noted $E_f$). This function based rule selection, whose formal definition is given in Figure 9, consists of finding the first rule that is made of a function whose application with the input string returns true.

If no such function based rule is found, then the lexer has to choose a regexp based rule.

Most lexers perform regexp based rule selection using a longest match selection policy based on the longest match and priority rules. That selection policy allows to select the first lexical rule whose regexp matches the longest prefix of the input string.

The Coqlex definition of this policy uses two concepts:

**prefix:** A string $p$ is said to be a prefix of a string $s$ if and only if there exists a string $s'$ such that $s = p ++ s'$. For examples $\epsilon$ is a prefix of any string.

**l-score:** Given a regexp $r$, a string $s$, and a natural number $n$, we say that the *l*-score of $r$ on $s$ is $n$, written as $\mathbb{S}_l(r, s) = n$, if and only if the length of the longest prefix of $s$ that $r$ can match is $n$. For example, the *l*-score of $[\![a]\!]^*$ in 'aabaaaa' is 2 as the longest prefix of 'aabaaaa' that $[\![a]\!]^*$ can match is 'aa' whose length is 2. There exist cases where there is no score (e.g: $\mathbb{S}_l([\![a]\!], 'bac')$). In that case, we note $\mathbb{S}_l(r, s) = -\infty$.

The inductive definition of our implementation of *l-score* computation is given in Figure 10.

To prove the correctness of $\mathbb{S}_l$, we used the Coq substring function of the Coq string module[21] to define the prefix. This function takes two natural numbers $n$ and $m$ and a string $s$, and returns the substring of length $m$ of $s$ that starts at position $n$ denoted by $\delta_n^m(s)$. Here, the position of the first character is 0. If $n$ is greater than the length $|s|$ of $s$ then $\epsilon$ is returned. If $m > (|s| - n)$, then $\delta_n^m(s) = \delta_n^{|s|-n}(s)$. Consequently, if $m \leq |s|$, then $\delta_0^m(s)$ is the prefix of length $m$ of $s$. For all strings $s$ and regexps $r$, we provided Coq proofs of the following theorems:

1. if there exists a natural number $n$ such that $\mathbb{S}_l(r, s) = n$, then $n \leq |s|$. This helps to make sure that $n$ can be used with $\delta$ to extract the prefix of length $n$.

$$\frac{}{E_l([],s) = \bot} \quad \frac{\mathbb{S}_l(r,s) = -\infty}{E_l((r,a)::t,s) = E_l(t,s)} \quad \frac{\mathbb{S}_l(r,s) = n \quad E_l(t,s) = (r_t, a_t, n_t) \quad n_t > n}{E_l((r,a)::t,s) = (r_t, a_t, n_t)}$$

$$\frac{\mathbb{S}_l(r,s) = n \quad E_l(t,s) = \bot}{E_l((r,a)::t,s) = (r,a,n)} \quad \frac{\mathbb{S}_l(r,s) = n \quad E_l(t,s) = (r_t, a_t, n_t) \quad n_t \leq n}{E_l((r,a)::t,s) = (r,a,n)}$$

**Figure 11** The formal description of the longest match selection policy. The symbol $\bot$ means that no rule is selected.

2. if there exists a natural number $n$ such that $\mathbb{S}_l(r,s) = n$, then $\delta_0^n(s) \in L(r)$. This means that the input regexp matches the prefix of length $n$ of $s$.
3. if there exists a natural number $n$ such that $\mathbb{S}_l(r,s) = n$, then for all $m$ such that $n < m \leq |s|$, $\delta_0^m(s) \notin L(r)$. This means that *l-score* is maximal. Therefore, there exists no prefix of length higher than $n$ that $r$ can match.
4. $\mathbb{S}_l(r,s) = -\infty$ if and only if for all natural number $m$, $\delta_0^m(s) \notin L(r)$.

Properties 1, 2, and 3 show that $\mathbb{S}_l$ is correct. This means that if a score is returned, this score is the length of the longest prefix of the input string that the input regexp can match. Property 4 shows the completeness and the soundness of $\mathbb{S}_l$. This means that if no score is found, then there is no score, and if there exists a score, $\mathbb{S}_l$ will return it.

Using $\mathbb{S}_l$, the longest match policy (noted $E_l$) consists in choosing the regexp based rule whose regexp has the highest *l-score*. The Coqlex formal definition of that policy is defined in Figure 11.

To prove the correctness of $E_l$, we proved with Coq that for every string $s$ and list $l_r$ of regexp based rules:

1. if $E_l(l_r, s) = \bot$ then for every regexp $r$ and semantic action $a$ such that $(r,a) \in l_r$, $\mathbb{S}_l(r,s) = -\infty$
2. if there exists a regexp $r$, a semantic action $a$ and a natural number $n$ such that $E_l(l_r, s) = (r, a, n)$ then for every regexp $r'$, semantic action $a'$ and natural number $n'$ such that $(r', a') \in l_r$ and $\mathbb{S}_l(r', s) = n'$, $n' \leq n$
3. for every regexps $r$ and $r'$, semantic actions $a$ and $a'$ and natural number $n$, if $E_l(l_r, s) = (r, a, n)$ and $\mathbb{S}_l(r', s) = n$ then $E_l((r', a')::l_r, s) = (r', a', n)$

Besides the longest match policy, Coqlex defines the shortest match policy that allows to select the first regexp based rules whose regexp matches the shortest prefix of the input string. The implementation technique of the shortest match policy is similar to the longest match policy. This implementation starts by the definition of the **s-score** (noted: $\mathbb{S}_s$) that allows to compute the length of the shortest prefix that a regexp can match. The formal definition of **s-score** is described in Figure 12.

Similarly to the $\mathbb{S}_l$, we proved the correctness and completeness of $\mathbb{S}_s$ through Coq proofs of the following theorems:

$$\frac{\text{nullable } r = \text{true}}{\mathbb{S}_s(r, \epsilon) = 0} \quad \frac{\text{nullable } r = \text{false}}{\mathbb{S}_s(r, \epsilon) = \infty} \quad \frac{\mathbb{S}_s(r/a, z) = n \quad \text{nullable } r = \text{false}}{\mathbb{S}_s(r, az) = n + 1}$$

$$\frac{\mathbb{S}_s(r/a, z) = \infty \quad \text{nullable } r = \text{false}}{\mathbb{S}_s(r, az) = \infty}$$

**Figure 12** The formal description of *s-score* computation.

$$\frac{}{E_s([], s) = \bot} \quad \frac{\mathbb{S}_s(r, s) = \infty}{E_s((r, a) :: t, s) = E_s(t, s)} \quad \frac{\mathbb{S}_s(r, s) = n \quad E_s(t, s) = (r_t, a_t, n_t) \quad n_t < n}{E_s((r, a) :: t, s) = (r_t, a_t, n_t)}$$

$$\frac{\mathbb{S}_s(r, s) = n \quad E_s(t, s) = \bot}{E_s((r, a) :: t, s) = (r, a, n)} \quad \frac{\mathbb{S}_s(r, s) = n \quad E_s(t, s) = (r_t, a_t, n_t) \quad n_t \geq n}{E_s((r, a) :: t, s) = (r, a, n)}$$

**Figure 13** The formal description of the shortest match selection policy.

1. if there exists a natural number $n$ such that $\mathbb{S}_s(r, s) = n$, then $n \leq |s|$. This helps to make sure that $n$ can be used with $\delta$ to extract the prefix of length $n$.
2. if there exists a natural number $n$ such that $\mathbb{S}_s(r, s) = n$, then $\delta_0^n \in L(r)$. This means that the input regexp matches the prefix of length $n$ of $s$.
3. if there exists a natural number $n$ such that $\mathbb{S}_s(r, s) = n$, then for all $m$ such that $m < n$, $\delta_0^m(s) \notin L(r)$. This means that *s-score* is minimal. Therefore, there exists no prefix of length lower than $n$ that $r$ can match.
4. $\mathbb{S}_s(r, s) = \infty$ if and only if for all natural number $m$, $\delta_0^m(s) \notin L(r)$.

Using $\mathbb{S}_s$, the shortest match policy consists in choosing the regexp based rule whose regexp has the lowest *s-score*. The Coqlex formal definition that policy is defined in Figure 13.

### 6.3 Coqlex Rule Selection

Using $E_f$, $E_l$ and $E_s$, the formal definition of the rule selection $E$ can be defined as follows:

$$E(E', l_r, l_f, s) = \begin{cases} E_f(l_f, s) \text{ if } E_f(l_f, s) \neq \bot \\ E'(l_r, s) \text{ otherwise} \end{cases}$$

where $E'$ is either $E_l$ or $E_s$. In the Coq code presented in Figure 1, $E$ is represented by generalizing_elector, $E_l$ is represented by longest_match_elector. The implementation of $E_s$ in the Coqlex library is represented by shortest_match_elector.

### 6.4 Optimization

The naive implementation suggested by the formal definition of $\mathbb{S}_l$ and $E_l$ has a time complexity that is at least quadratic in the size of the input string. In fact, the implementation of *l-score* requires reading all the characters of the input string for every regexp based lexical rule and thus for every token. However, this is not necessary in some cases (e.g for all string $s$ with $\mathbb{S}_l(\varnothing_r, s) = -\infty$).

To increase the performance of $\mathbb{S}_l$, $\mathbb{S}_s$, $E_l$ and $E_s$, we implemented a regexp simplification function which is based on the following properties:

**The alternative:** $r + \varnothing_r \equiv r$ and $\varnothing_r + r \equiv r$

**The concatenation:** $r \cdot \varnothing_r \equiv \varnothing_r$, $\varnothing_r \cdot r \equiv \varnothing_r$, $r \cdot \epsilon_r \equiv r$, $\epsilon_r \cdot r \equiv r$ and $r^* \cdot r^* \equiv r^*$

**The Kleene star:** $\varnothing_r^* \equiv \epsilon_r$, $(r^*)^* \equiv r^*$ and $\epsilon_r^* \equiv \epsilon_r$

**The difference:** $r - \varnothing_r \equiv r$ and $\varnothing_r - r \equiv \varnothing_r$

These simplifications aim to detect when a given regexp is equivalent to a regexp whose score is trivial (e.g $\varnothing_r$ or $\epsilon_r$). We proved these properties in Coq and then used the smart constructor technique[8] to write an optimized version of the regexp derivative function. That function works similarly to the original one, except that it returns a simplified version of the derivative. Then, we also rewrote the **s-score** and **l-score** functions to use the optimized version of the regexp derivative function and return the result for trivial cases. For example, we proved that for every $r$ and $s$,

$$\mathbb{S}_s(r^*, s) = \mathbb{S}_s(\epsilon_r, s) = \mathbb{S}_l(\epsilon_r, s) = 0 \qquad \mathbb{S}_s(\varnothing_r, s) = \infty \qquad \mathbb{S}_l(\varnothing_r, s) = -\infty$$

We proved that the optimized **s-score** and **l-score** are equal to the original ones. We propagated this optimization to $E_l$ and $E_s$ and obtained performance in linear time in the size of the input string (see Section 7 below). In fact, during lexical analysis, the number of tokens is high, and the number of symbol reads required to produce each of them can be low. Before those optimizations, the *l-score* computation required reading the whole input string to produce a token, whereas in the optimized version we stop reading when the score is trivial, typically when the input string is equivalent to $\varnothing_r$ or $\epsilon_r$. This allows to reduce drastically the number of character reads and so the execution time.

For performance reasons, we modified the **s-score** and **l-score** functions to return the lexeme and the remaining string during score computation. We also proved that the modified function is correct.

## 7 Evaluation: Syntax and Expressivity

We are now going to compare Coqlex with OCamllex (4.14.0), the OCaml standard lexer generator, Flex (2.6.4), the C standard lexer generator, and Verbatim++ (github Dec. 2021, https://github.com/egolf-cs/Verbatim), the current state of the art for lexers verified in Coq. It is common for Coq users to implement unverified part of their programs in OCaml, so we can expect that Coqlex would replace an existing OCamllex lexer; we also compare with flex, which is very similar to OCamllex in practice, to show that standard unverified lexer generators form a meaningful common baseline.

We compare the convenience of the user-facing syntax in which the lexer is described, and the expressivity of its semantics – which lexing rules can or cannot be easily expressed in the system.

Our broad claim is that using Coqlex is very similar to using traditional lexer generators such as OCamllex, while Verbatim++ has a noticeably more verbose syntax and is less expressive.

### 7.1 Surface Syntax Example

To evaluate the convenience of using each system, we display a fragment of the JSON lexers in Figure 16 in the Appendix. (We wrote the flex, OCamllex and Coqlex versions; the Verbatim++ lexer was written by the Verbatim++ authors.)

Without going into the details of all systems here, we claim that Coqlex is closer in convenience to flex or OCamllex than to Verbatim++.

In regard to the syntax of the .vl files, the *Coqlex generator* is built to process a language that is very close to flex or OCamllex. This means that there are only few differences between .vl files and their equivalent .l or .mll files. For example, the listing on the right in Figure 1 presents the OCamllex equivalent of the Coqlex lexer presented in the left of that Figure.

### 7.2 Labels

As seen in the Verbatim++ example in Figure 16, Verbatim++ lexers use a notion of *labels*, a data-type returned after the selection. Semantic actions are functions that take that label and lexeme to return a token. Therefore, semantic actions do not have access to the remaining string and thus, cannot perform recursive calls. A consequence of this is that Verbatim++ lexers cannot ignore parts of the input string (such as comments and extra spaces): all input characters must be included in a lexeme.

### 7.3 Looping Lexers

Coqlex lets the user express potentially-nonterminating lexers, using the "fuel" technique to express non-termination inside Coq. This is not as powerful as requiring termination proofs, but we believe that it is more practical. It does at least let users consume a variable number of input characters (not always strictly more than one) and call a lexing rule recursively, which is a common occurrence in real-world lexers.

We originally required each rule to consume at least one character, but found out that it several limits expressivity compared to OCamllex: in those situations, allowing recursive calls is *good* for usability.

In this section, we discuss what happens in the *bad* case of a lexer with a termination bug due to recursive calls.

Let us consider the Coqlex lexer and the OCamllex lexer specified in Figure 14. Regarding those specifications, the lexers are supposed to work as follows:

```
1  (**Coqlex**)
2
3  rule my_lexer = parse
4    | 'b' 'a'* 'b' { ret 0 }
5    | 'a'* { my_lexer }
6    | EOF { ret 1 }
```

```
1  (**OCamllex**)
2
3  rule my_lexer = parse
4    'b' 'a'* 'b'  { 0 }
5    | 'a'*    { my_lexer lexbuf }
6    | EOF     { 1 }
```

**Figure 14** An example of a lexer whose execution can loop defined using Coqlex and OCamllex syntax

- If the remaining string is $\epsilon$ then 1 is returned.
- Else if the longest prefix of the input string in lexbuf matches $[[b]] \cdot [[a]]^* \cdot [[b]]$ then 0 is returned
- Else if it matches $[[a]]^*$ it performs a recursive call on the remaining string of lexbuf (updated after the selection). This is a common technique used to ignore elements such as comments during lexical analysis.

When such lexer is called with a string $s$ that starts with a character that is different from 'a' and 'b', the selection chooses the semantic action that is associated with the regex $[[a]]^*$ and which score is 0. This means that the lexeme is $\epsilon$ and the remaining string is $s$. As the semantic action associated with this regexp is a recursive call, it leads to an infinite loop.

The behavior of each of the tools is as follows:
- The lexer generated by OCamllex (and flex) from the code on the right in Figure 14 loops when the input string is 'c'.
- The lexer generated by Coqlex from the code on the left in Figure 14 runs out of fuel and returns an error to the user (in finite time).
- Verbatim++ does not handle this kind of problem because semantic actions do not allow recursive calls.

### 7.4 Proofs about the Generated Lexers

Finally, the simplicity of the Coqlex lexer specifications allows to write proofs on Coqlex lexers. For example, we have proven that the looping lexer defined in the listing on the left in Figure 14 always returns an error related to the fuel when the first character of the input string is different from 'a' and 'b'.

## 8 Evaluation: Performance

The implementation approach of Coqlex is very simple, we use Brzozowski derivatives, with the derivation operations performed on the fly during lexing. In contrast, standard lexer generators use a sophisticated compilation step to finite automata [3] for fast regexps matching. Verbatim++ also advertises a DFA compilation step. As a result, our initial expectation would be that Coqlex-generated lexers are noticeably slower than standard lexers and Verbatim++-generated lexers.

**Coqlex: Generating Formally Verified Lexers**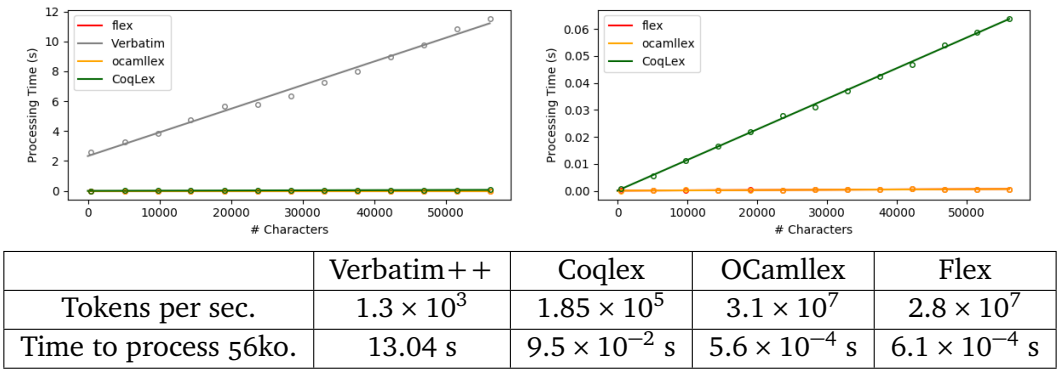

|  | Verbatim++ | Coqlex | OCamllex | Flex |
|---|---|---|---|---|
| Tokens per sec. | $1.3 \times 10^3$ | $1.85 \times 10^5$ | $3.1 \times 10^7$ | $2.8 \times 10^7$ |
| Time to process 56ko. | 13.04 s | $9.5 \times 10^{-2}$ s | $5.6 \times 10^{-4}$ s | $6.1 \times 10^{-4}$ s |

■ **Figure 15** Comparison of execution time in seconds for Coqlex, OCamllex, flex and Verbatim++ lexers on Verbatim++ JSON benchmark.

Note that the typical goal for verified tools is not to match non-verified tools in performance – this is rarely realistic – but rather to provide "good enough" performance for the tool use-cases.

We evaluated the execution time of the generated lexical analysers in two phases. For the first phase, we evaluated their performance on United States GDP data from the past several decades[2], stored in JSON format, the same benchmark used to evaluate the performances of Verbatim++.

We started with analysing a JSON lexer implemented by the Verbatim++ developers using Verbatim++ Coq source code, then we used Coqlex, flex and OCamllex generators to generate lexers with very similar specifications. We compared the time performance in Figure 15; we notice three groups separated by two order of magnitudes each, ocamllex and flex which are the fastest, Coqlex in the middle, and Verbatim++ as the slowest tool.

We can see from the results that Verbatim++ starts with a constant-time cost of 2s, independently of the input size. This is due to the compilation phase to DFAs that is performed on each run of the lexer. Even if we ignore this constant-time overhead, Verbatim++ is about two orders of magnitude slower (130x) than Coqlex, which is itself about two orders of magnitude slower than OCamllex (156x) or flex (75x).

We also evaluated the performance of an XML lexer across Verbatim++, Coqlex and OCamllex, on the sample data from [15] and obtained similar performance results.

**Is this too slow?** Coqlex is two orders of magnitude slower than unverified lexer generators. Is this usable in practice, or much too slow? In our experience, Coqlex meets the "good enough" bar in terms of performance for use-cases we care about. In a typical compiler, lexing and parsing combined take a small fraction of the compilation time; slowing this part a lot has a noticeable but acceptable effect on compile times.

For example, in our industrial use-case for Coqlex presented in Section 4, we wrote two variants of our compiler, one using a verified frontend (Coqlex + verified parser), and one using an unverified frontend. On the largest compiled project (25MB of source code), the compiler with unverified frontend takes around 2 minutes, and the compiler with the verified front takes 6 minutes. It is unusual to spend such a large fraction of compile time in the frontend, but "good enough": the resulting compile times are perfectly acceptable to our users.

Note that the target audience for Coqlex is not all language-processing tools, but *verified* compilers. (There is little point in verifying only the frontend.) The typical users of verified compilers are typically willing to sacrifice some performance in exchange for safety guarantees, and they often already have heavy-weight code-production processes due to verification or certification requirements. In our industrial use-case, the tool running in 2 or 6 minutes is part of a much more complex industrial toolchain that, on the 25MB project we mentioned, takes 5 hours in total.

**Why is Verbatim++ so slow?** Our work was not originally built on Verbatim or Verbatim++, but an independent project done at around the same time. When we discovered Verbatim++ we expected its more elaborate implementation and focus on optimization to get much better performance than Coqlex. We studied the Verbatim++ implementation and try to propose explanations for the performance difference. For reasons of space, this discussion is moved to the Appendix B.

# 9 Conclusion

Coqlex proposes a Coq library that implements all the common features used in lexical analysis and provides a Coq proof of their correctness. It also brings a Code generator that reads a lex-like lexer description and outputs verified lexers written in Coq using the Coqlex library. By design, Coqlex is built to be usable and simple. The Coqlex library has been used to build the lexer of Coqlex generator. The Coqlex generator and library have been used in the implementation of a formally verified optimizing compiler. Despite its simplicity, lexers generated by the Coqlex generator or implemented using the Coqlex library have good enough performance, noticeably better than the existing tool Verbatim++.

Even if the execution time performance of Coqlex can be improved further, it lays strong foundations for verified lexer generation.

As an alternative to unverified lexer generators, it allows to make one more step in proving end-to-end correctness of compilers. It has already found applications in the real world, with an industrial use-case for a verified lexer for a subset of Ada.

## 9.1 Related Work

The formal correctness of lexing does not seem to be extensively studied in the literature. For instance, even for the formally proven compiler CompCert [14], lexing is one of the phases which are not formally verified.

In existing approaches to verify lexers, like in CakeML [12], the lexer is implemented by hand (without using a generator) and proven equal to a simple and deterministic function. Most lexers are more complicated and it can be hard to find a simple and deterministic function that is equal to the lexer. Nipkow[17] formally verifies a regex-to-DFA translation, but to our knowledge, this work was not packaged as a user-oriented executable tool.

**9.2 Future Work**

Some performance improvements could come from extraction to native OCaml strings (or arrays), or the memoization of the derivation function using axiomatized imperative data structures. To improve performance further, we could translate lexing rules to a proper DFA representation. There are two independent questions:

**How to do the computation.** We should take inspiration from the excellent description of DFA construction in [18]. They introduce the notion of "derivative classes" to compute equivalence classes of characters for a given redex, and reduce the number of transitions. They also introduce the notion of "regex vector" to derive all lexing rules in parallel, instead of having a separate automata construction for each rule. However, there is a real possibility that the size of the resulting automaton would be too big for Coq to handle in practice. Unverified lexer generators use sophisticated tricks to compress the description of the automata, which could be difficult to verify, and may not even suffice to ensure practicality inside Coq. If the approach of generating deterministic automata proved impractical, we could instead generate a non-deterministic automata, for example using Antimirov derivatives [1], which can be sensibly more compact at the cost of a reasonable performance overhead.

**When to do the computation.** Verbatim++ starts computing the lexing automata whenever the user wants to get the lexing function from the lexing rules, which is typically whenever the Coq-implemented user program is invoked. In our experiment, the JSON benchmark for example needs two seconds of automata computation before they start accepting input.

A convenient lexer generator should avoid this constant overhead by generating the automaton during a meta-programming phase ahead of the actual lexing, but doing this in a verified setting is not obvious. We played with the Verbatim++ implementation and tried asking Coq to do the computation at type-checking time, but the Coq virtual machine fails in practice as the computation requirements are too large. It may be possible that a more efficient automata construction, using derivative classes or an NFA construction, could be done at type-checking time in this way. Another approach would be to do the automata construction at Coqlex preprocessing time, and only have Coq validate the correspondence of the lexing rules and the automata. This requires either producing a Coq proof script during this metaprogramming step, or defining a Coq tactic that is flexible enough to validate all such correspondences.

To improve the usability of Coqlex, we would like to allow users to conveniently bind substrings matched by regexps. Coqlex users can prove explicitly that their lexer is terminating, but it would be nice if Coqlex could generate a termination argument in simple cases, or assist in the termination proof even in subtle cases. Finally, implementing a .mll to .vl converter would ease adoption for frontends already using OCamllex.

## A   Example: JSON lexers in Flex, OCamllex, Coqlex, and Verbatim++

```
/**flex**/

int_re  "-"?("0"|([1-9][0-9]*))
%%
{ws_carr_re}  { return WS; }
{int_re}      { yylval.i = atoi(yytext); return INT; }
{float_re}    { yylval.f = atof(yytext); return FLOAT; }
"true"   { return TRUE; }
"false"  { return FALSE; }
"null"   { return NULL; }
%%
```

```
(**OCamllex**)

let int_re = '-'? ('0' | (['1'-'9'] ['0'-'9']*))
rule read =
  parse
  | ws_carr_re   { WS }
  | int_re
    { INT (int_of_string(Lexing.lexeme lexbuf)) }
  | float_re
    { FLOAT (float_of_string(Lexing.lexeme lexbuf)) }
  | "true"  { TRUE } | "false"  { FALSE }
  | "null"  { NULL }
```

```
(**Coqlex**)

let int_re = '-'? ('0' | (['1'-'9'] ['0'-'9']*))
rule read =
  parse
  | ws_carr_re   { ret WS }
  | int_re
    { ret_l (fun str => INT (mint_of_string str)) }
  | float_re
    { ret_l (fun str => FLOAT (mfloat_of_string str)) }
  | "true"  { ret TRUE } | "false"  { ret FALSE }
  | "null"  { ret NULL }
```

```
(**Verbatim**)

(* semantic actions *)
Definition apply_sem (pr : Label * String) :=
match pr with
| (INT, z) => match (String2int z) with
  | Some i => Some (existT sem_ty INT i)
  | None => None
  end
| (FLOAT, z) => match (String2int z) with
  | Some i => Some (existT sem_ty FLOAT i)
  | None => None
  end
| (STRING, z) =>
  Some (existT _ STRING (string_of_list_ascii z))
| (L, _)   => Some (existT _ L tt)
end.

[...]

(* regex *)
Definition digit_re := stringUnion "0123456789".
Definition nz_digit_re := stringUnion "123456789".
Definition pos_re := App nz_digit_re (Star digit_re).
Definition zero_re := stringApp "0".
Definition nat_re := Union zero_re pos_re.
Definition int_re :=
  App (Optional (stringApp "-")) nat_re.

Definition read := lex_sem [
  (WS, ws_carr_re); (INT, int_re);
  (FLOAT, float_re);
  (TRUE, stringApp "true");
  (FALSE, stringApp "false");
  (NULL, stringApp "null") ].
```

**Figure 16**   Comparing JSON lexer for OCamllex, Flex, Coqlex and Verbatim++

## B  Why is Verbatim++ so Slow?

In a regexp matcher based on Brzozowski's derivatives, we compute a derivative of the "current" regexp each time we read a character from the input. This derivation function dominates the running time of the matcher. Optimizations to derivate-based lexers typically try to optimize the derivation function or cache its computation.

A naively-written lexer using derivatives is quadratic, for two different reasons:

1. The size of the "current" regexp may grow on each successive derivation. This occurs when using the naive/texbook definition of derivation, typically on the kleene star. (In the worst case, the size increase due to derivation may even lead to an exponential blowup, but this worst case is typically not encountered in real-world regexps.)

   The standard approach to avoid this size increase, and the ensuing quadratic behavior, is the use of *smart constructors*, which perform simplifications on the fly during derivation and keep the regexp size constant in common cases.

2. It is important to stop regexp matching as early as possible, when the derived regexp becomes empty. A naive implementation may keep deriving the empty regexp until the end of the input string is reached. If each call to the lexer traverses the whole input to produce one token, we get quadratic behavior.

In Coqlex, being careful of these two issues sufficed to produce a reasonably efficient regexp matcher. (We experimented with memoizing the derivation function, which gives a further 2x speedup, but is hard to do in a fully-verified way as it typically involves unverified imperative data structures.)

Our hypothesis is that Verbatim made more complex implementation choices to avoid the quadratic blowup that do result in linear lexing performance but with very large book-keeping overhead:

1. In a precomputation phase, they compute a very large set of possible repeated derivatives, stored in an associative data structure that Verbatim++ calls a DFA: the states of the automaton are regexps. (This corresponds to the constant 2s overhead observed on the JSON grammar.)

2. The precomputed table of derivates is represented by a trie. Looking up a derivative requires navigating the trie, running comparison functions between current regexp and the regexps in the trie.

3. On top of this caching layer, Verbatim++ memoizes the score for each pair (current regexp, input position). This removes the second source of quadratic behavior (stopping early on failure) in an indirect way: the very first score computation will traverse the whole input with an empty (failed) regexp, in linear time, but any further call that reaches the empty regexp will find the (empty regexp, current position) pair in the table and stop early.

The composition of these two layers (DFA represented as a trie + score memoization) runs in linear time, but with considerable constant-factor overheads.

We wondered if the simpler optimization approach that we used in Coqlex would suffice to give good performance to Verbatim++ — is it "just" missing smart constructors or stop-early logic? So we modified the Verbatim++ derivation function to use smart constructors as we presented in Section 6.4; we call this variant *VerbatimSC*. We also modified the Verbatim++ score computation function to stop early when the input regexp is $\emptyset_r$ (there is no match) or $\epsilon_r$ (score is 0); we call this variant *VerbatimSt*.

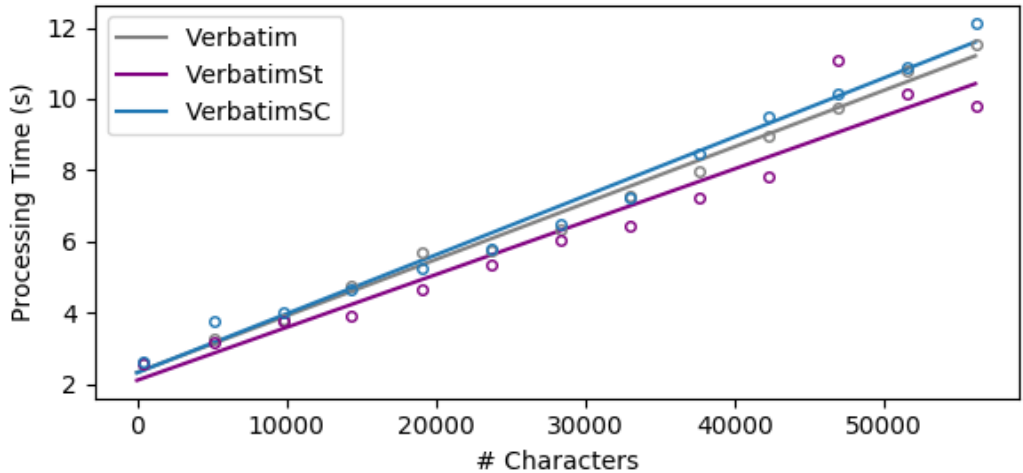

**Figure 17** Comparison of execution time in seconds for Verbatim++, VerbatimSC and VerbatimSt on Verbatim++ JSON benchmark.

The performance of these two variants are compared to Verbatim++ in Figure 17. *VerbatimSC* is slower than Verbatim++: Verbatim++ already performs simplifications during the DFA computation, and our smart constructors only add additional overhead. On the other hand, (*VerbatimSt* and its simple stop-early logic does provide a significant improvement to Verbatim++. Unfortunately, it still remains sensibly slower than Coqlex.

Comparing the design of Coqlex and Verbatim++, we also notice some smaller design choices that have a performance impact:

1. Better regexp language. Coqlex extended its regex definition (see Section 6.1) to reduce the execution in usual cases; for example, the interval [a-z] can be represented and checked as a single construct, while Verbatim++ uses a large sum requiring 26 tests in the worst case.
2. Better data representation after extraction. Coqlex uses OCaml native ascii characters while Verbatim++ uses the default Coq representation of ASCII characters as 8-uplets of booleans, which is at least 8x slower, probably more.
3. The Coqlex score computation function has been designed to return the score, the lexeme and the remaining string at the same time, while Verbatim++ first computes the lexeme and remaining string and then computes the score (the length of the lexeme) to chose the longest.

## C  The Generated Code of OCamllex for Mini-cal

```
 1  (*OCamllex input*)
 2
 3  {
 4  open Lexing
 5  open TokenDefinition
 6  }
 7  let ident = ['a'-'z']+
 8  let numb = ['0'-'9']+
 9
10  rule minlexer = parse
11  | '\n' { new_line lexbuf; minlexer lexbuf }
12  | ident {ID (Lexing.lexeme lexbuf)}
13  | numb { Number (Lexing.lexeme lexbuf)}
14  | '+' { PLUS }
15  | '-' { MINUS }
16  | '*' { TIMES }
17  | '(' { LPAREN }
18  | ')' { RPAREN }
19  | eof { Eof }
20  | _ { failwith ("unknown token : " ^ (
          ↪  Lexing.lexeme lexbuf))}
21
22  (* trailer section *)
23  {}
```

```
 1  (*OCamllex output*)
 2
 3  open Lexing
 4  open TokenDefinition
 5
 6  let __ocaml_lex_tables = {
 7    Lexing.lex_base = "\000\000\246 ...";
 8    Lexing.lex_backtrk = "\255\255...";
 9    Lexing.lex_default = "\001\000...";
10    Lexing.lex_trans = "\000\000...";
11    Lexing.lex_check = "\255\255...";
12    Lexing.lex_base_code = "";
13    Lexing.lex_backtrk_code = "";
14    Lexing.lex_default_code = "";
15    Lexing.lex_trans_code = "";
16    Lexing.lex_check_code = "";
17    Lexing.lex_code = "";
18  }
19
20  let rec minlexer lexbuf =
21      __ocaml_lex_minlexer_rec lexbuf 0
22  and __ocaml_lex_minlexer_rec lexbuf __ocaml_lex_state =
23  match Lexing.engine __ocaml_lex_tables __ocaml_lex_state
          ↪  lexbuf with
24  | 0 -> ( new_line lexbuf; minlexer lexbuf )
25  | 1 -> (ID (Lexing.lexeme lexbuf))
26  | 2 -> ( Number (Lexing.lexeme lexbuf))
27  | 3 -> ( PLUS )
28  | 4 -> ( MINUS )
29  | 5 -> ( TIMES )
30  | 6 -> ( LPAREN )
31  | 7 -> ( RPAREN )
32  | 8 -> ( Eof )
33  | 9 -> ( failwith ("unknown token : " ^ (Lexing.lexeme lexbuf)))
34  | __ocaml_lex_state -> lexbuf.Lexing.refill_buff lexbuf;
35      __ocaml_lex_minlexer_rec lexbuf __ocaml_lex_state;;
```

**Figure 18** Comparing OCamllex input and output for mini-cal

## D  Source Code Organization

The Coqlex source code is organized as follows:

**The directory regexp_opt:** contains the implementation of Coqlex extended version of regular expression based on Brzozowski derivatives.

**RValues.v:** contains the definition of usual regex (string, character, identifiers, numbers...)

**RegexpSimpl.v** contains the implementation of regexp simplification described in Section 6.4.

**MachLen.v:** contains the implementation of the score used to perform the longest match rule. Similarly, ShortestLen.v contains the score function of the shortest match rule.

**MachLenSimpl.v:** contains the optimized version of the longest match rule score computation. Similarly, ShortestLenSimpl.v contains the optimized score function of the shortest match rule.

**LexerDefinition.v:** contains the Coqlex selection system and data-type definition.

**SubLexeme.v:** contains the Coqlex sub-lexeme functions and proofs.

**CoqlexUtils.v:** contains the definition of usual semantic action.

**CoqlexLexer.v:** contains definition of the lexer of the *Coqlex generator*.

**Extraction.v:** contains the extraction directives of the .v files above.

**Parser.vy:** contains definition of the parser of the *Coqlex generator*.

**coqlex.ml:** contains *Coqlex generator* main function.

**ParserUtils.ml and LexerUtils.ml:** contains OCaml function that facilitates the use of the OCaml extracted code of Coqlex lexers.

**The directory example:** contains examples of Coqlex lexer specified by .vl files, their OCamllex equivalent and benchmark data.

**The directory Comparison:** contains JSON and XML benchmark data, Verbatim++, OCamllex and Coqlex lexers and a python code that allowed to plot the Figure 15.

## About the authors

**Wendlasida Ouedraogo** PhD student.
wendlasida.ouedraogo@lix.polytechnique.fr
https://orcid.org/0000-0002-5331-4121

**Gabriel Scherer** Research scientist.
https://gallium.inria.fr/~scherer/
https://orcid.org/0000-0003-1758-3938

**Lutz Straßburger** Senior research scientist.
http://www.lix.polytechnique.fr/Labo/Lutz.Strassburger/
https://orcid.org/0000-0003-4661-6540